\documentclass[twocolumn]{fairmeta}
\usepackage{wrapfig}
\usepackage{tabularx}
\usepackage{textcomp}
\usepackage{stfloats}
\usepackage{url}
\usepackage{verbatim}
\usepackage{graphicx}
\usepackage{titlesec}
\usepackage{tocloft}
\usepackage{adjustbox}
\usepackage{enumitem}
\usepackage{multirow}
\usepackage{pifont}
\usepackage{tikz}
\usepackage{comment}
\usepackage{amsmath,amssymb} % define this before the line numbering.
\usepackage{colortbl}  %彩色表格需要加载的宏包
\usepackage{color}
\usepackage{booktabs} 
\usepackage{hyperref}
\usepackage{graphicx}    % 用于插入图片
\usepackage{subcaption} 
\usepackage{booktabs}
\usepackage{amsmath} % 数学公式支持
\usepackage{multirow} % 表格多行合并
\usepackage{booktabs} % 表格更好的线条
\usepackage{subcaption} % 支持 subtable 和 subfigure
\usepackage{todonotes}
\RequirePackage{xspace}
\makeatletter
\DeclareRobustCommand\onedot{\futurelet\@let@token\@onedot}
\def\@onedot{\ifx\@let@token.\else.\null\fi\xspace}

\makeatother

\definecolor{adptorange}{RGB}{248, 205, 172}
\definecolor{cmpblue}{RGB}{189, 215, 238}
\definecolor{cmpblue}{RGB}{189, 215, 238}

\definecolor{our_red}{RGB}{232,157,160}
\definecolor{our_blue}{RGB}{136,206,230}
\definecolor{our_orange}{RGB}{246,200,168}
\definecolor{our_green}{RGB}{178,211,164}

\definecolor{attn_code0}{RGB}{247,215,200}
\definecolor{attn_code1}{RGB}{238,169,139}
\definecolor{mlp_code0}{RGB}{204,201,221}
\definecolor{mlp_code1}{RGB}{102,95,153}

\definecolor{token_blue}{RGB}{84, 120, 140}
\usepackage{pifont}       % \ding{xx}
\usepackage{bbding}       % \Checkmark,\XSolid,... (需要和pifont宏包共同使用)
\usepackage{fontawesome}
\usepackage{xspace}

\usepackage{float}

\newlength\savewidth

\newcolumntype{x}[1]{>{\centering\arraybackslash}p{#1pt}}
\newcolumntype{y}[1]{>{\raggedright\arraybackslash}p{#1pt}}
\newcolumntype{z}[1]{>{\raggedleft\arraybackslash}p{#1pt}}

\renewcommand{\paragraph}[1]{\vspace{1mm}\noindent\textbf{#1}}
\usepackage{colortbl}
\usepackage{xcolor}
\usepackage{wrapfig}

\renewcommand{\paragraph}[1]{\vspace{1.25mm}\noindent\textbf{#1}}

\usepackage{algorithm}
\usepackage{listings}

\definecolor{codeblue}{rgb}{0.25, 0.5, 0.5}
\definecolor{codekw}{rgb}{0.35, 0.35, 0.75}
\lstdefinestyle{Pytorch}{
    language = Python,
    backgroundcolor = \color{white},
    basicstyle = \fontsize{9pt}{8pt}\selectfont\ttfamily\bfseries,
    columns = fullflexible,
    aboveskip=1pt,
    belowskip=1pt,
    breaklines = true,
    captionpos = b,
    commentstyle = \color{codeblue},
    keywordstyle = \color{codekw},
}

\definecolor{green}{HTML}{009000}
\definecolor{red}{HTML}{ea4335}

\title{MoDA: Multi-modal Diffusion Architecture for Talking Head Generation}

\author{%
  Xinyang Li$^{1,2}$, 
  Gen Li $^{2}$,
  Zhihui Lin$^{1,3}$, 
  Yichen Qian$^{\dagger1,3}$,
  Gongxin Yao$^{2}$,
  Weinan Jia$^{1}$, 
  Aowen Wang$^{1,2}$, 
  Weihua Chen$^{1,3}$,
  Fan Wang$^{1,3}$\\
  \small{
  $^1$Xunguang Team, DAMO Academy, Alibaba Group, 
  $^2$Zhejiang University, 
  $^3$Hupan Lab
  }\\
  \small{$^\dagger$ Corresponding author.}\\
  \texttt{yichen.qyc@alibaba-inc.com, l\_xyang@zju.edu.cn}
}

\abstract{
Talking head generation with arbitrary identities and speech audio remains a crucial problem in the realm of the virtual metaverse. 
Despite progress, current methods still struggle to synthesize diverse facial expressions and natural head movements while generating synchronized lip movements with the audio.
The main challenge is stylistic discrepancies between speech audio, individual identity, and portrait dynamics. 
To address the challenge of inter-modal inconsistency, we introduce MoDA, a multi-modal diffusion architecture with two well-designed technologies. First, MoDA explicitly models the interaction among motion, audio, and auxiliary conditions, enhancing overall facial expressions and head dynamics. In addition, a coarse-to-fine fusion strategy is employed to progressively integrate different conditions, ensuring effective feature fusion. Experimental results demonstrate that MoDA improves video diversity, realism, and efficiency, making it suitable for real-world applications. 
Project Page: https://lixinyyang.github.io/MoDA.github.io/}

\date{\today}
\usepackage{multirow}
\usepackage{graphicx}
\begin{document}
\maketitle
\begin{figure*}
  \centering
  \includegraphics[width=1\linewidth]{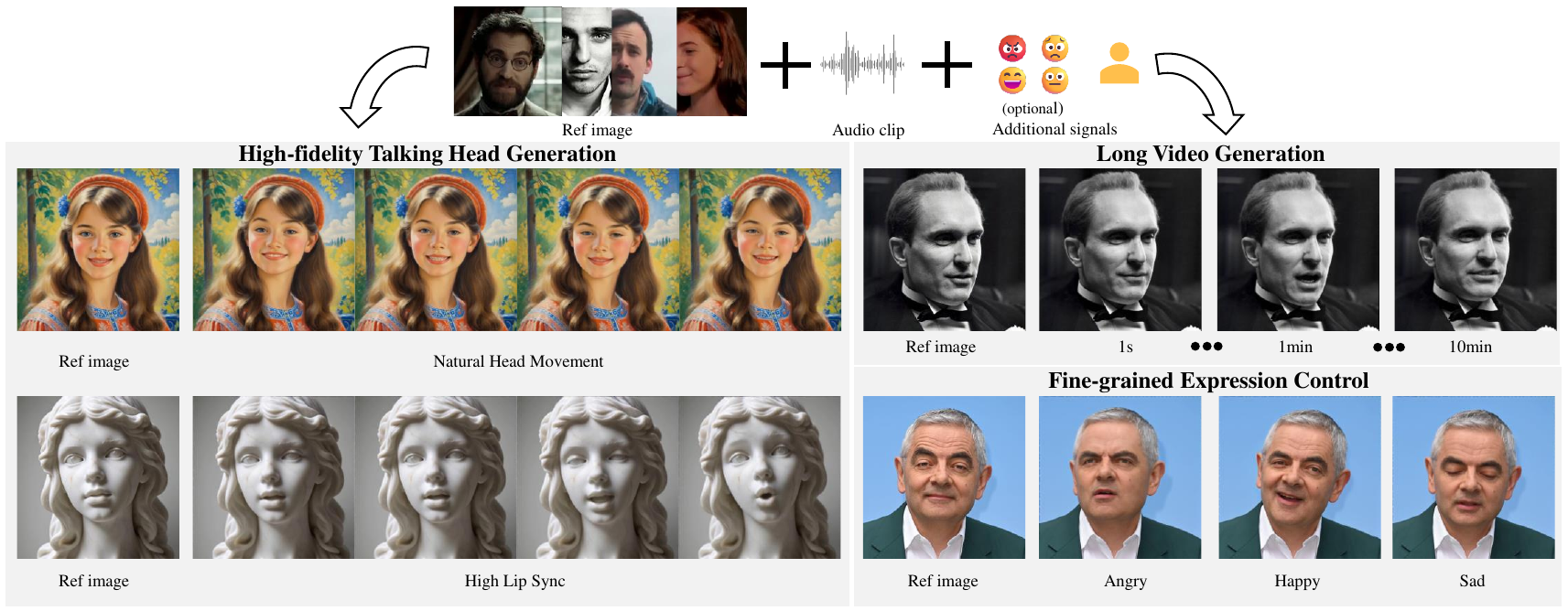}
  \caption{Given a reference image, an audio clip, and other control signals, MoDA excels in producing vivid portrait animation videos in real-time.  Beyond basic lip synchronization, it demonstrates strong capabilities in generating diverse facial expressions and natural head movements. Furthermore, it supports fine-grained facial expression control and long-term video synthesis.}
  \label{fig:teaser}
\end{figure*}

\section{Introduction}
Talking head generation aims to create a photorealistic, speaking portrait from a single image, guided by audio and other modalities. Combined with the generative adversarial network (GAN)~\cite{goodfellow2014generative} and diffusion
model~\cite{Sohl-Dickstein_Weiss_Maheswaranathan_Ganguli_2015}, recent methods demonstrate
widespread potential applications, such as immersive telepresence, virtual characters, and augmented reality.  

\begin{figure}
    \centering
    \includegraphics[width=1\linewidth]{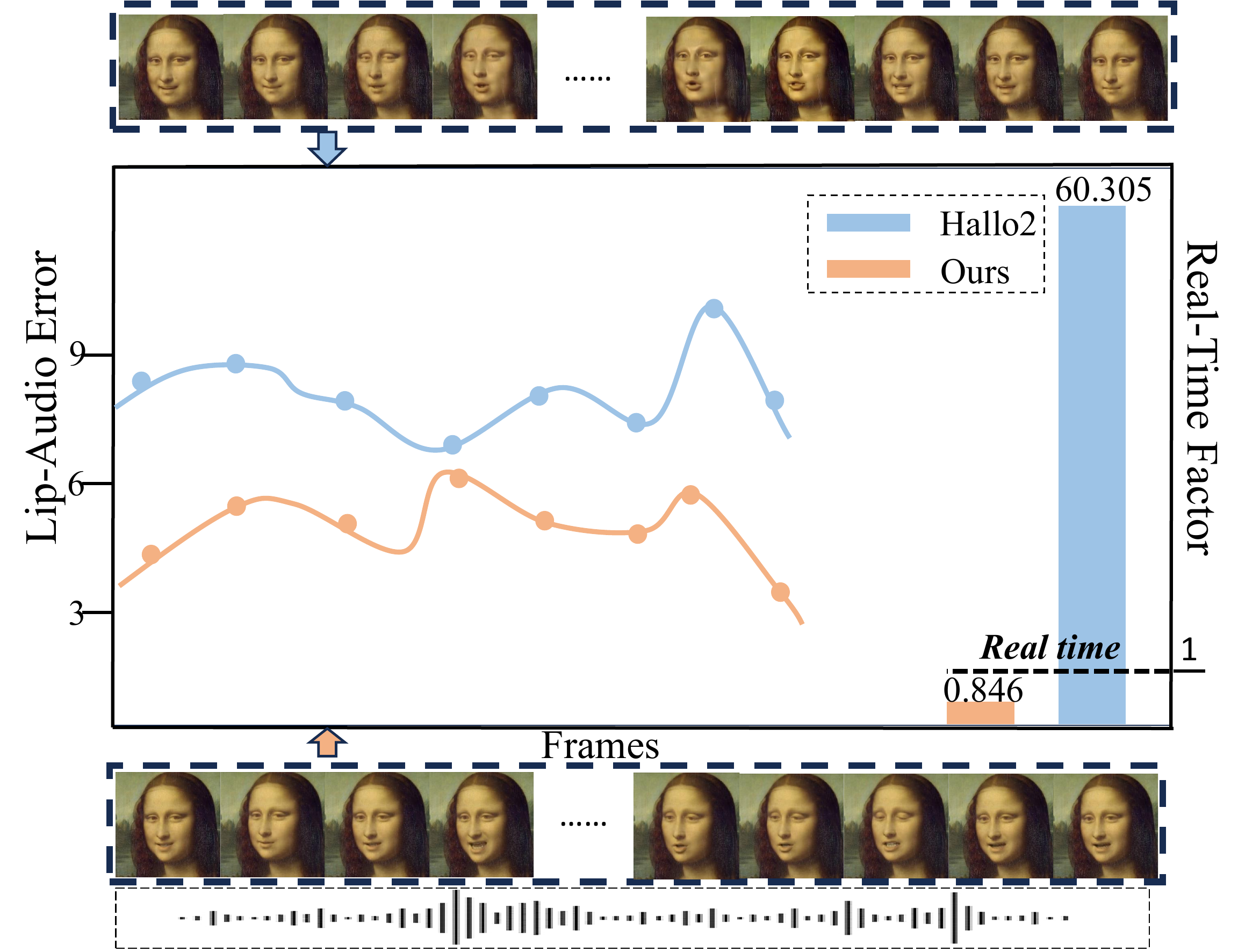}
    \caption{Comparison of our approach with hallo2~\cite{cui2024hallo2}. Real-Time Factor (RTF) measures the ratio of inference time to output duration. An RTF <1 indicates that the system can perform real-time generation. Frame-by-frame comparison. Hallo2 shows lower lip synchronization, inference efficiency, and identity consistency than our method.}
    \label{fig:compare}
\end{figure}

Diffusion models have recently marked a significant advancement in generative modeling, enabling the creation of highly diverse videos. Early diffusion-based methods~\cite{cui2024hallo2, tian2024emo, jiang2024loopy, xu2024hallo} generate the final video directly from the audio input.  Although trainable from end to end, methods like Hallo2~\cite{cui2024hallo2} remain two major limitations persist, as shown in Fig.~\ref{fig:compare}: 1) Inefficient inference process and visual artifacts. 2) Unnatural facial expressions and head movements with precise lip-sync. Recently, two-stage methods~\cite{xu2024vasa, cao2024joyvasa, li2024ditto} have simplified the diffusion process by bypassing complex variational auto-encoder (VAE) decoding. Methods like VASA-1~\cite{xu2024vasa} first use the diffusion model to generate intermediate motion representations from audio, and then use a separate rendering network to synthesize the final video. However, the final video quality is heavily dependent on the accuracy of these intermediate representations. Thus, these methods still struggle to achieve natural facial dynamics with precise lip-sync due to suboptimal predictions.

Looking into the aforementioned issues, we argue that their root cause is the stylistic discrepancies between speech audio, individual identity, and portrait dynamics. These methods typically concatenate multiple conditions to form a mixed representation, which is then fed into a cross-attention mechanism where information flows only from this mixed modality to motion. This design introduces a learning bias, causing the model to focus only on the most shallow feature cues while neglecting the intricate relationships between the modalities. Consequently, it fails to handle more complex or conflicting scenarios. As in the ablation study, this limitation leads directly to inconsistent motion sequences when the model is conditioned on arbitrary identity input.

To address the challenge of inter-modal inconsistency, we propose MoDA, a novel framework designed for synergistic talking head generation. MoDA begins by operating within a joint parameter space that bridges motion generation and neural rendering, encompassing a disentangled motion-appearance space~\cite{guo2024liveportrait} along with audio, emotion, and identity. This space has a dimensionality that is an order of magnitude lower than traditional VAE spaces, which dramatically reduces the complexity of multi-modal fusion. Moreover, by incorporating optional conditions like identity and emotion, MoDA makes the generative modeling of complex distribution more tractable and increases fine-grained control over the generative process.

MoDA is guided by two core principles designed to address these inconsistencies: 1) We draw inspiration from recent lip-to-speech tasks~\cite{9956600, prajwal2020learning}, where visual information can provide additional context to complement the audio. As shown in Fig.~\ref{fig:liuchengtu1}, MoDA introduces the Multi-modal Diffusion Transformer (MMDiT)~\cite{esser2024scaling}, equipped with rectified flow~\cite{liu2022flow}, as the framework to facilitate multi-modal fusion. In this design, MoDA can dynamically adapt audio features based on motion, identity, and emotion, thereby improving the accuracy of motion generation. 2) To systematically integrate these modalities based on semantic information, MoDA implements a coarse-to-fine fusion strategy. Initially, the model uses separate weights to capture the unique characteristics of each modality. 
In the intermediate stage, a unified representational space is introduced for semantically linked modalities like audio, emotion, and identity to form a unified motion command.
Finally, all modalities are integrated into a unified representation space, allowing holistic fusion. To further encourage precise lip-sync while maintaining motion diversity, MoDA provides an optional Adaptive Lip-motion sync Expert (ALSE), which can be integrated during training. 
The contributions can be summarized as follows:
\begin{itemize}
\item This paper proposes MoDA, a novel multi-modal diffusion framework that generates high-fidelity talking head videos from an image, audio, and additional modalities. 
\item A coarse-to-fine fusion strategy is designed to progressively integrate noisy motion with audio and other modalities, enabling effective multi-modal fusion.
\item Extensive evaluations on public datasets demonstrate that our method outperforms contemporary alternatives in visual quality and quantitative metrics.
\end{itemize}
\section{Related Work}
\subsection{Disentangled Face Representation}
In recent years, extensive research has been conducted on disentangled facial representation learning. Some methods utilize sparse keypoints~\cite{siarohin2019first, zakharov2020fast} or 3D Morphable Model (3DMM)~\cite{Blanz_Vetter_1999, li2017learning} techniques to explicitly represent facial dynamics. The 3DMM initially projects the 3D head shape into several low-dimensional Principal Component Analysis (PCA) spaces, which provide orthogonal bases. Based on these orthogonal bases, attributes such as identity, pose, and expression can be manipulated in the 3D model using linear blend skinning (LBS). However, these methods may face issues such as inaccurate reconstructions or limited ability to decouple facial attributes. Recent learning-based approaches, such as Face Vid2Vid~\cite{Wang_Mallya_Liu_2021}, LivePortrait~\cite{guo2024liveportrait}, and Megaportrait~\cite{drobyshev2022megaportraits}, have been introduced to disentangle facial representations within a nonlinear parameter space, significantly enhancing the expressive power of these models. Moreover, by leveraging a more compact parameterization, these methods can effectively capture intricate details and dynamic facial expressions, offering greater flexibility in both facial reconstruction and animation.
\subsection{Audio-driven Talking Head Generation}
In audio-driven digital human technology, one-shot methods have gained considerable attention, enabling the generation of dynamic and expressive avatars from a single image input. Existing methods can generally be classified into two categories: single-stage and two-stage audio-to-video generation. The former~\cite{jiang2024loopy, cui2024hallo2, prajwal2020lip, shi2024joyhallo, xu2024hallo, guan2023stylesync, cui2024hallo3} directly maps audio features to video frames in an end-to-end manner. In contrast, the latter\cite{li2024ditto, xu2024vasa, cao2024joyvasa, sun2023vividtalk, zhang2023sadtalker} introduces an intermediate representation, such as motion sequences or keypoints, serving as a bridge between audio and video synthesis.

Early single-stage methods~\cite{guan2023stylesync, prajwal2020lip, suwajanakorn2017synthesizing} leveraged GANs to focus on lip-sync accuracy while keeping other facial attributes unchanged. Recent advancements have incorporated diffusion-based approaches~\cite{xu2024hallo, shi2024joyhallo, cui2024hallo2, tian2024emo, chen2024echomimic}, expanding the scope of research by mapping audio inputs to diverse facial expressions and continuous natural head movements. However, these methods rely on denoising and rendering images within the VAE space, where appearance and motion remain highly entangled, resulting in substantial computational overhead and low inference efficiency. 

The two-stage methods address the above limitations by introducing a disentangled facial space as an intermediate representation. Early methods~\cite{ye2023geneface, ye2022audio} used explicit landmarks or 3DMM as intermediate representations to bridge the gap between audio inputs and animated outputs, allowing for more controllable and interpretable motion synthesis. Recently, VASA-1~\cite{xu2024vasa}, Ditto~\cite{li2024ditto}, and JoyVASA~\cite{cao2024joyvasa} have shifted towards implicit facial representations~\cite{Wang_Mallya_Liu_2021, drobyshev2022megaportraits, guo2024liveportrait} while employing DIT-based models for audio-to-motion mapping, leading to more expressive and natural video synthesis. However, relying solely on cross-attention for lip-sync generation neglects the rich multimodal interactions and deep-level information within the input signals, thereby limiting the diversity and expressiveness of the generated outputs.
\begin{figure*}
    \centering
    \includegraphics[width=1.0\linewidth]{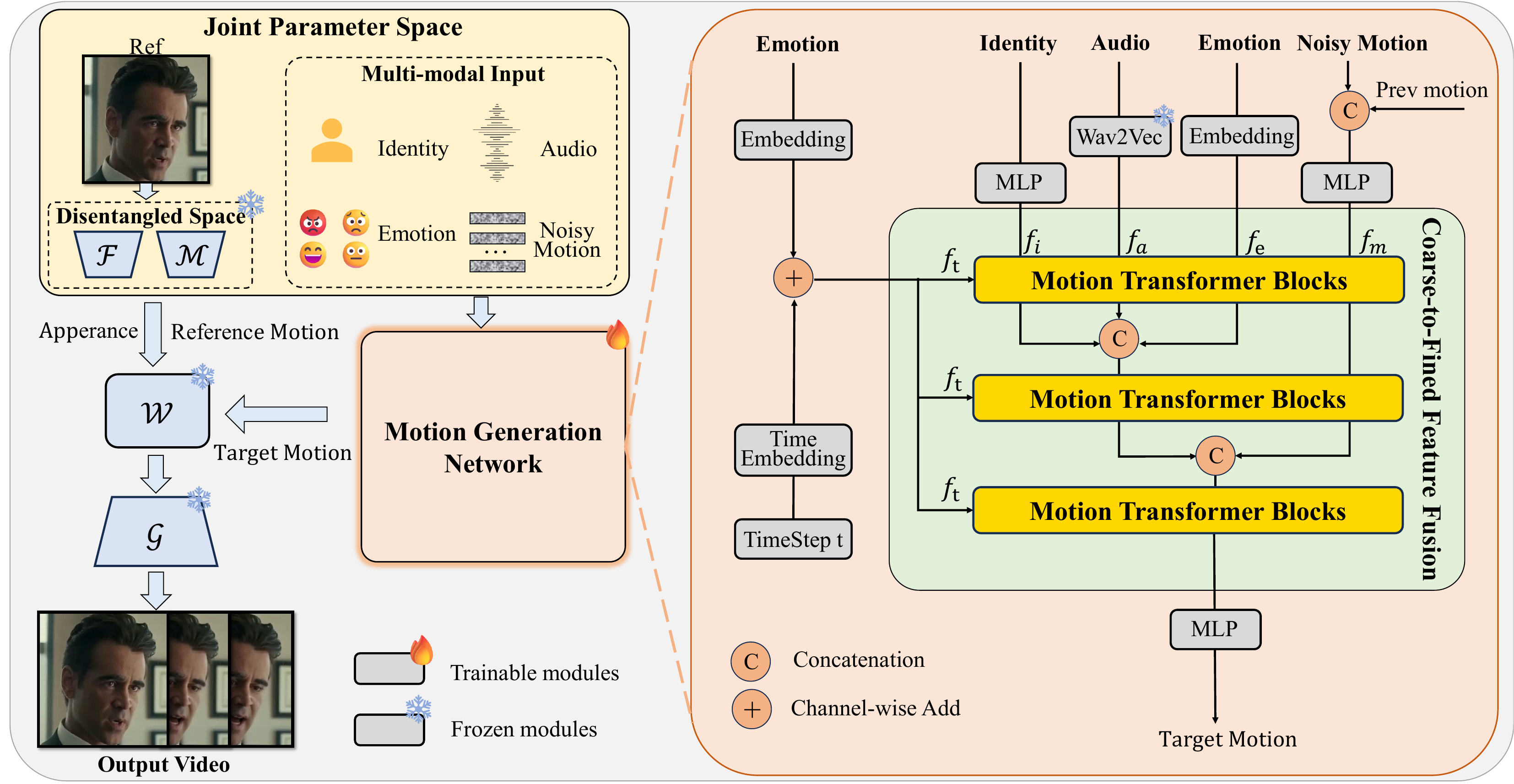}
    \caption{Overall architecture of the proposed MoDA and illustration of our Motion Generation Network. The appearance extractor $\mathcal{F}$, motion extractor $\mathcal{M}$, warping module $\mathcal{W}$, and decoder $\mathcal{G}$ are frozen. A motion feature generation model based on a Diffusion Transformer is then trained to generate motion features.}
    \label{fig:liuchengtu1}
\end{figure*}
\section{Method}
\subsection{Preliminaries}\label{sec:3.1}
Denoising Diffusion probabilistic models (DDPMs)~\cite{song2020denoising} have emerged as a powerful framework for generative modeling by formulating the data generation process as an iterative denoising procedure. In the forward diffusion process, Gaussian noise $\epsilon$ is gradually introduced into the data distribution across $T$ discrete timesteps, producing noisy latent features:$z_t = \sqrt{\alpha_t} z_0 + \sqrt{1 - \alpha_t} \epsilon,$ where $\alpha_t$ represents a variance schedule that determines the noise level at each timestep, and $z_0$ is the raw data. The model is trained to reverse this process by taking the noisy latent representation $z_t$ as input and estimating the added noise $\epsilon$. The training objective is defined as:$\mathcal{L} = \mathbb{E}_{z_t, c, \epsilon \sim \mathcal{N}(0,1), t} \left[ || \epsilon - \epsilon_\theta (z_t, t, c) ||_2^2 \right],$ where $\epsilon_\theta$ denotes the noise prediction generated by the model, and $c$ represents additional conditioning signals, such as audio or motion frames, which are particularly relevant in the generation of talking videos.

Recently, Stable Diffusion 3 (SD3)~\cite{esser2024scaling} has advanced the paradigm by introducing Rectified Flow~\cite{pooladian2023multisample, liu2022flow} that optimizes the traditional DDPMs objective:
\begin{equation}
\mathcal{L} = \mathbb{E}_{z_t, c, \epsilon \sim \mathcal{N}(0,1), t} \left[ ||(\epsilon-z_0) - v_\theta (z_t, t, c) ||_2^2 \right],
    \label{eq:2}
\end{equation}
where $z_t = (1-t) z_0 + t \epsilon$, and $v_\theta$ denotes the velocity field. After the rectified flow training is completed, the transition from $\epsilon$ to $z_0$ can be formulated using the numerical integration of an ordinary differential equation (ODE):
\begin{equation}
z_{t - \frac{1}{N}} = z_t + \frac{1}{N} v_\theta (z_t, t, c),
    \label{eq:rectified_flow}
\end{equation}
where $N$ denotes the discretization steps of the interval $[0,1]$. The piecewise linear denoising process improves training stability. Furthermore, since our audio-to-motion task does not involve complex pixel information, this approach is particularly well-suited to our needs. Given these advantages, we adopt the rectified flow for training.
\subsection{Model Architecture}\label{sec:3.2}
Instead of simply concatenating these conditioning signals, we enhance the intrinsic characteristics and emotional nuances of the audio representation by treating external emotion and identity cues as "catalysts". This approach balances identity and emotion between audio and speakers, enabling more natural lip control in real-world scenarios.
\subsubsection{Joint Parameter Space}\label{sec:3.2.1}
As shown in Fig.~\ref{fig:liuchengtu1}, we incorporate the existing facial re-enactment framework~\cite{guo2024liveportrait} to extract disentangled facial representations. Specifically, the motion extractor $\mathcal{M}$ yields expression deformations $\delta$, head pose parameters $(R, t)$, the canonical keypoints of the source image $x_{c}$, and a scaling factor $S$. The motion representation $(R_s, \delta_s, t_s, S_s) \in \mathbb{R}^{70}$, serves as an identity-agnostic representation of the source input and is used to train MoDA to predict $(\hat{R}, \hat{\delta}, \hat{t}, \hat{S})$ given audio input. 
\begin{equation}
\begin{alignedat}{2}
x_s &= S_s \cdot (x_c R_s + \delta_s) + t_s &\quad&, \\
\hat{x} &= \hat{S} \cdot (x_c \hat{R} + \hat{\delta}) + \hat{t} &\quad&.
\end{alignedat}
\end{equation}
Subsequently, the warping field estimator $\mathcal{W}$ computes a field from $x_s$ and $\hat{x}$ to deform the 3D features $f_s$, which are then passed to the generator $\mathcal{G}$ to synthesize the target image.
The audio features $f_a$ are extracted using the wav2vec~\cite{Schneider_Baevski_Collobert_Auli_2019} encoder. To maintain consistency in the identity feature space across various scenarios, we use the canonical keypoints of the source image $x_c$ as identity information and generate identity features $f_i$. For facial emotion signals, we use a visual emotion classifier~\cite{Savchenko_2022} to extract the speaker's emotional labels and encode them into corresponding features $f_e$. The motion features from the first frame of each clip are used as guiding conditions to ensure inter-frame continuity and generate noisy motion features $f_m$. Inspired by EMO2~\cite{tian2025emo2}, emotion features $f_e$ are added to timestep embeddings to generate timestep features $f_t$, which are injected into each motion transformer block by adaptive layer normalization (AdaLN)~\cite{Peebles_Xie_2022}. AdaLN is used to prevent the degradation of emotion features during the joint-attention operation, ensuring that emotional cues are preserved throughout the fusion. By integrating these conditioning signals, our model effectively generates realistic and temporally consistent motion features.
\begin{figure}
    \centering
    \includegraphics[width=1\linewidth]{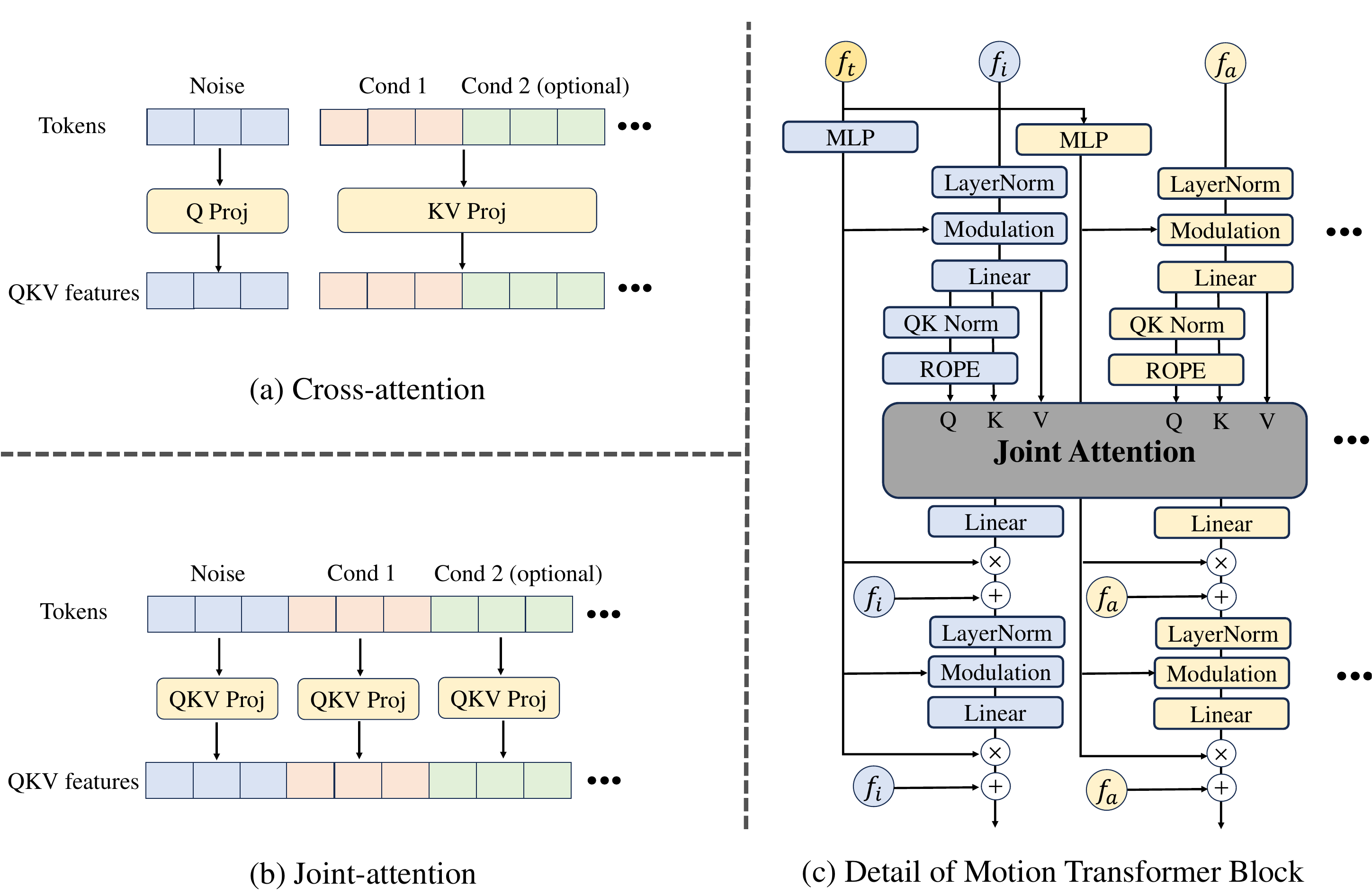}
    \caption{The detail of the Motion Transformer Block and the Joint-attention. (a) Cross-attention uses noise as the query and conditions as key and value. (b) Joint-attention projects noise and conditions separately and concatenates them before attention.}
    \label{fig:liuchengtu2}
\end{figure}

\subsubsection{Motion Transformer Blocks}\label{sec:3.2.2}
As shown in Fig.~\ref{fig:liuchengtu2} (c), the Motion Transformer Block consists of three main components: Modality-specific paths, Joint-attention and Rotational Position Encoding.

Modality-specific paths are designed to distinguish representations from different modalities. Specifically, each modality is equipped with its own adaLN and a modulation mechanism~\cite{Peebles_Xie_2022} to improve the conditional generation capabilities of the model.

Joint-attention is employed to enable interaction across different modalities. All modalities are first projected onto their respective query (Q), key (K), and value (V) representations, which are then concatenated in order along the sequence dimension. The combined sequence is processed through an attention operation, after which the attended features are split back into their respective modalities in the original order.

To enhance temporal alignment between noisy motion and other conditional features, we adopt Rotational Positional Encoding (RoPE)~\cite{yang2024cogvideox} instead of the absolute positional encoding used in MMDiT. Specifically, we first expand the expression and identity features to match the sequence length of the noisy motion and audio features. We then apply aligned RoPE across all modalities, which facilitates better temporal synchronization and more consistent feature representations.
\subsubsection{Coarse-to-Fined Feature Fusion }\label{sec:3.2.3}
Although assigning modality-specific paths with separate QKV projections and FeedForward Networks (FFNs) in the attention mechanism can enhance multi-modal information, this design often overlooks the inherent semantic commonalities shared across modalities. Independently learned weights hinder effective multi-modal feature fusion and introduce redundant parameters, potentially causing inconsistency issues. 
This issue is particularly pronounced in tasks like audio-to-motion generation, where no pixel-level inputs are involved, and the semantic gap between modalities is relatively small. In such cases, maintaining entirely separate parameterizations fails to provide meaningful benefits and instead leads to duplicated representations.
To enhance the model's capacity for multi-modal understanding, we introduce a Coarse-to-Fine Feature Fusion strategy. This fusion strategy progressively integrates multi-modal features while reducing unnecessary parameters, thereby enhancing the multi-modal understanding and training stability.

As illustrated in Fig.~\ref{fig:liuchengtu1}, the architecture progresses through three stages: a four-stream, two-stream, and single-stream process.
In the four-stream stage, each modality is processed independently to learn modality-specific representations, facilitating early-stage feature differentiation.
In the two-stream stage, the audio stream is concatenated with emotion and identity features to form a merged stream that shares weights. This design encourages a balanced integration of emotional and identity cues from both the audio and the reference image.
Finally, all modalities are unified into a single representation stream to allow deeper fusion and enhance generative expressiveness.

\begin{table*}[h]
\caption{Comparison with existing methods on the HDTF and CelebV-HQ test sets. ↑ Higher is better. ↓ Lower is better. Best results are in \textbf{bold}, second-best are \underline{underlined}.}
\centering
\setlength{\tabcolsep}{6pt} % Adjusted column spacing
\small
\begin{tabular}{l|lcccccc} % Added '|' after 'l' for Dataset and Method
\toprule
\textbf{Dataset} & \textbf{Method} 
& \textbf{FVD} $\downarrow$ 
& \textbf{FID} $\downarrow$  
& \textbf{F-SIM} $\uparrow$ 
& \textbf{Sync-C} $\uparrow$ 
& \textbf{Sync-D} $\downarrow$  
& \textbf{Smo(\%)} $\uparrow$ \\
\midrule
\multirow{8}{*}{\textbf{HDTF}} 
&GT           & -        & -        & 0.860 & 7.267 & 7.586 & 0.9959 \\

&EchoMimic    & 207.987  & \underline{29.633} & 0.887 & 2.744 & 11.805 & 0.9939 \\
&JoyHallo     & 256.226  & 44.842   & 0.852 & 7.360 & 7.984  & 0.9944 \\
&Hallo        & 216.573  & 34.350   & 0.878 & 7.087 & 7.941  & 0.9950 \\
&Hallo2       & 229.806  & 34.426   & 0.871 & 7.102 & 7.976  & 0.9951 \\
&Ditto        & 243.491  & 32.200   &  \underline{0.943} &6.102 & 8.790  & 0.9970 \\
&JoyVASA      & 229.634  & 32.584   &  \textbf{0.953} & 5.255 & 9.600  & 0.9968 \\
&\textbf{Ours (sync)}  & \underline{191.292} & 30.449 & 0.925 & \textbf{8.183} & \textbf{7.065} & \underline{0.9970} \\
&\textbf{Ours}         & \textbf{174.622}  & \textbf{28.182} & 0.927 & \underline{7.369} & \underline{7.744} & \textbf{0.9971} \\
\midrule
\multirow{8}{*}{\textbf{CelebV-HQ}} 
&GT           & -        & -        & 0.861 & 5.837 & 7.989 & 0.9964 \\
&EchoMimic    & 258.451  & 47.169   & 0.837 & 2.610 & 11.216 & 0.9946 \\
&JoyHallo     & 282.081  & 57.247   & 0.813 & \underline{6.041} & 8.418 & 0.9945 \\
&Hallo        & 245.101  & 44.411   & 0.851 & 5.629 & 8.384 & 0.9952 \\
&Hallo2       & 242.352  & 46.615   & 0.851 & 5.671 & 8.397 & 0.9953 \\
&Ditto        & 302.525  & 46.996   & 0.915& 4.681 & 9.280 & \textbf{0.9973} \\
&JoyVASA      & 271.231  & 44.574   & \textbf{0.918} & 5.171 & 8.632  & 0.9971 \\
&\textbf{Ours (sync)}  & \textbf{205.307} & \underline{44.201} & \underline{0.916} & \textbf{6.552} & \textbf{7.635} & \underline{0.9972} \\
&\textbf{Ours}         & \underline{205.442} & \textbf{44.071} & 0.913 & 5.878 & \underline{8.135} & 0.9972 \\
\bottomrule
\end{tabular}
\label{tab:comparison1}
\end{table*}
\subsubsection{Loss Function}\label{sec:3.2.4}
Our loss function utilizes the rectified flow loss and Eq.~ref{eq:2} is rewritten as:
\begin{equation}
\mathcal{L}_{RF} = \mathbb{E}_{z_t, c, \epsilon \sim \mathcal{N}(0,1), t} \left[ ||x - v_\theta (z_t, t, c) ||_2^2 \right],
    \label{eq:4}
\end{equation}
where $x$ represents $\epsilon-z_0$, $v_\theta$ denotes the velocity field. The velocity loss $L_{vel}$ is introduced to encourage improved temporal consistency:
\begin{equation}
\mathcal{L}_{vel} = ||x' - m' ||_2^2+||x'' - m'' ||_2^2 ,
    \label{eq:5}
\end{equation}
where $m$ denotes the output of $v_\theta (z_t, t, c)$, and $m''$, $m''$ denote the first-order and second-order derivatives of $m$. To further enforce lip-sync accuracy, we design the ALSE pretrained in audio and motion features and compute the loss $\mathcal{L}_{ALSE}$. It is worth noting that supervision from the pretrained ALSE is optional and can be applied as needed. In summary, the final loss can be expressed as follows:
\begin{alignat}{2}
\mathcal{L} &= \mathcal{L}_{RF} + \mathcal{L}_{vel} + \lambda_{sync} \cdot \mathcal{L}_{ALSE} &\quad&, \label{eq:5a} \\
\lambda_{sync} &= 
\begin{cases}
1, & \text{if} \ \mathcal{L}_{ALSE} < \tau \\
0, & \text{otherwise}
\end{cases}
&\quad&, \label{eq:5b}
\end{alignat}
where $\tau$ is set to 0.4 and controls when the synchronization loss is activated, ensuring that lip-motion supervision is applied only after basic motion diversity has been established. Further details about the Adaptive Lip-motion Synchronization Expert (ALSE) can be found in the supplementary materials.
\subsection{Realtime Inference}
The real-time conversational scenario is enabled by low-latency motion generation. We align audio features with the video frame rate and segment the audio into continuous 100-frame chunks for streaming generation. Additionally, leveraging low-dimensional intermediate representations and the rectified flow, we reduce DiT inference denoising steps from 50 to 10, yet achieve even higher quality. For detailed comparison, please refer to the supplementary material.
\subsubsection{Classifier-free guidance (CFG)}
In the training stage, we randomly assign each of the input conditions, and during inference, we perform the following:
\begin{equation}
\hat{z}^0 = (1 + \sum_{c \in C} \lambda_c) \cdot v_\theta(z^t, t, C) - \sum_{c \in C} \lambda_c \cdot v_\theta(z^t, t, C | c = \emptyset),
    \label{eq:6}
\end{equation}
where $\lambda_c$ is the CFG scale of condition c. $C | c = \emptyset$ denotes that the condition c is $\emptyset$.
\begin{figure*}
    \centering
    \includegraphics[width=0.9\linewidth]{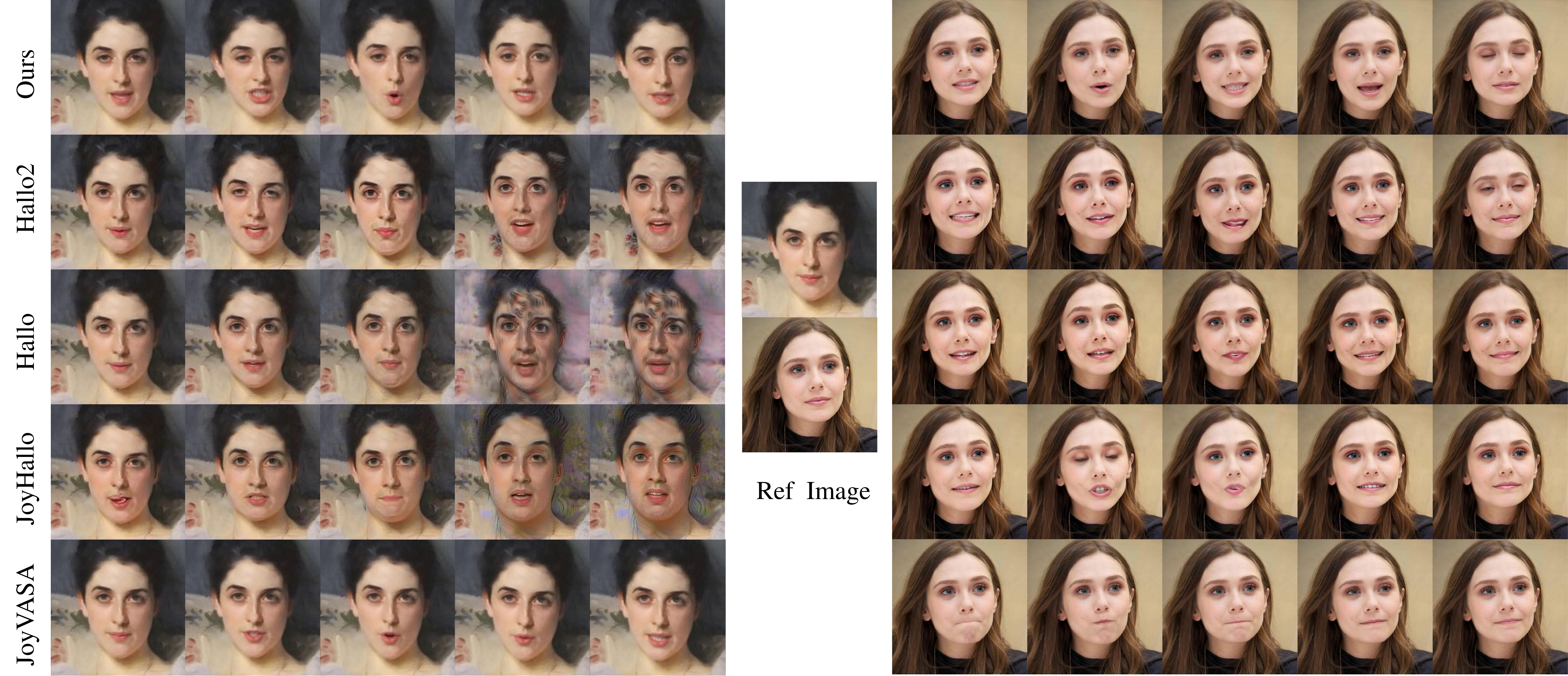}
    \caption{Qualitative Comparisons with State-of-the-Art Method. As single-frame images cannot fully represent synchronization, naturalness, and stability, we provide complete video comparisons in the supplementary materials.}
    \label{fig:shiyantu1}
\end{figure*}
\section{Experiments}
\subsection{Experiment Settings}
\subsubsection{Dataset and Metrics}
MoDA is primarily trained on three publicly available datasets: HDTF~\cite{zhang2021flow}, CelebV-Text~\cite{yu2023celebv}, and MEAD~\cite{wang2020mead}. For evaluation, we conducted experiments on three distinct test sets. The first two are derived from public datasets, CelebV-HQ~\cite{zhu2022celebv} and HDTF, each consisting of 50 randomly sampled clips ranging from 3 to 10 seconds in length. The third is an in-the-wild set with 20 diverse cases, including real individuals, animated characters, dynamic scenes, and complex headwear. Each sample is paired with audio that is speech, emotional dialogue, or singing. We utilize several evaluation metrics to assess the performance of the proposed method. The Fréchet Inception Distance (FID)~\cite{seitzer2020pytorch} and the Fréchet Video Distance (FVD)~\cite{unterthiner2019fvd} are used to assess the quality of the generated output, while the F-SIM~\cite{tian2024emo} measures facial similarity. In addition, Sync-C~\cite{chung2017out} and Sync-D~\cite{chung2017out} metrics are introduced to evaluate lip-synchronization between different methods. A temporal smoothness metric (Smo)~\cite{huang2024vbench} is also utilized to monitor the continuity of generated motion.
\subsubsection{Implementation Details}
During training, we randomly sample 80-frame segments from video clips to train the motion generation model. The model is trained for approximately 500 epochs on 8 NVIDIA H20 GPUs with a batch size of 512, using the Adam optimizer with a learning rate of 1e-4. During training, we apply a dropout probability of 0.1 for each emotion condition, while the dropout probability for speech is set to 0.5. Furthermore, the model is structured with 3 four-stream blocks, 6 two-stream blocks, and 12 single-stream blocks to progressively enhance multi-modal feature integration.
\begin{table}[h]
\centering
\setlength{\tabcolsep}{3pt} 
\caption{Comparison with existing methods on the wild test dataset. The best results
are bold, and the second are underlined.}
\begin{tabular}{lcccc}
\toprule
\textbf{Method}  
& \textbf{F-SIM} $\uparrow$ 
& \textbf{Sync-C} $\uparrow$ 
& \textbf{Sync-D} $\downarrow$  
& \textbf{RTF} $\downarrow$ \\
\midrule
EchoMimic    
& 0.870
& 2.292
& 12.130 
& 48.657  \\
JoyHallo   
& 0.825
& \underline{7.000} 
& 8.167 
& 59.735 \\
Hallo      
& 0.848 
& 6.051 
& 8.730 
& 59.190 \\
Hallo2     
& 0.849 
& 6.386 
& 8.523 
& 60.305 \\
% \midrule
Ditto   
& \underline{0.923}
& 6.107
& 9.040
& \textbf{0.792} \\
JoyVASA    
& \textbf{0.924}
& 5.569
& 9.368
& 1.717 \\
\textbf{Ours (sync)}      
& 0.896
& \textbf{7.710} 
& \textbf{7.469} 
& \underline{0.846} \\
\textbf{Ours}        
& 0.895
& 6.862 
& \underline{8.088} 
& \underline{0.846}\\
\bottomrule
\end{tabular}
\label{tab:comparison2}
\end{table}
\subsection{Results and Analysis}
We juxtapose the results of the proposed method against those of EchoMimic~\cite{chen2024echomimic}, Ditto~\cite{li2024ditto}, JoyVASA~\cite{cao2024joyvasa}, JoyHallo~\cite{shi2024joyhallo}, Hallo~\cite{xu2024hallo}, and Hallo2~\cite{cui2024hallo2}, Ours and Ours (Sync) (MoDA with ALSE).
\subsubsection{Quantitative Comparison}
The quantitative results on the CelebV-HQ test and HDTF dataset are shown in Table~\ref{tab:comparison1}. On the HDTF dataset, MoDA consistently outperforms all existing methods across all evaluation metrics. In particular, our method achieves the lowest FID and FVD scores, outperforming the second-best methods by 4.9 \% and 16.0 \%, respectively. This demonstrates the superiority of our method in terms of visual naturalness and the overall quality of the generated frames. The two-stage methods better preserve identity, as evidenced by their F-SIM scores. Notably, JoyVASA and Ditto achieve the highest and second-highest F-SIM, respectively—likely due to their relatively constrained head movements and facial expressions, which enhance frame-wise structural similarity. The correspondingly lower FVD scores further reflect these limitations. In contrast, MoDA effectively mitigates such issues, demonstrating that enhancing multi-modal fusion can lead to improved overall model performance.

On the CelebV-HQ test dataset, MoDA consistently outperforms the six baseline methods in all metrics except Sync-D and Smo, highlighting its robustness. Although MoDA slightly underperforms JoyHallo in synchronization confidence (Sync-C), it achieves a notable improvement in synchronization distance (Sync-D), along with further gains in motion smoothness. Tables~\ref{tab:comparison1} also reports the results with and without ALSE. Introducing ALSE to supervise the synchronization between audio and keypoints leads to a significant improvement in lip-sync.

Table~\ref{tab:comparison2} shows that our method outperforms existing approaches across multiple metrics on the wild test set with diverse identities. We also report the real-time factor (RTF), where RTF \textless 1 indicates real-time capability. Ditto achieves the lowest RTF, benefiting from TensorRT acceleration.  Our method also delivers competitive efficiency, demonstrating its suitability for real-world applications.
\begin{figure*}
    \centering
    \includegraphics[width=0.9\linewidth]{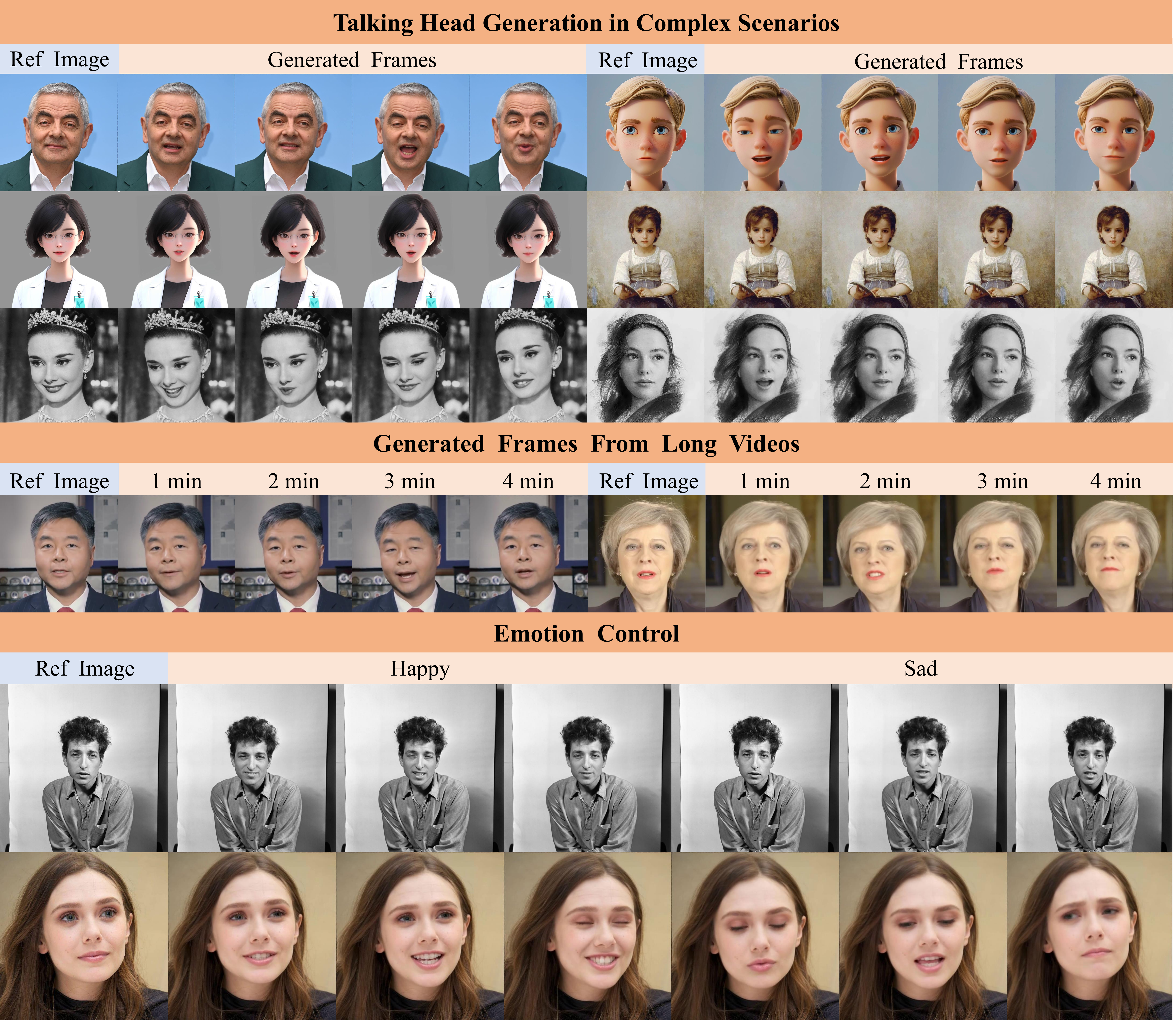}
    \caption{Generation results for portraits and audio in diverse styles are presented. We also demonstrate long-video inference and fine-grained control over facial expressions.}
    \label{fig:shiyantu2}
\end{figure*}
\subsubsection{Qualitative Evaluation} \label{sec:4.2.2}
As shown in Fig.~\ref{fig:shiyantu1}, we selected two types of cases from the wild dataset for visual comparison. For each character, we generated videos using each method and selected frames from the same location for comparison. Analyzing the results, previous one-stage-based methods, including Hallo2, Hallo, and JoyHallo, suffer from appearance blurring and unnatural expressions during temporal inference due to the strong entanglement between appearance and motion. In contrast, our proposed two-stage framework effectively mitigates these issues, ensuring high consistency in the generated details. Compared to the two-stage method JoyVASA, MoDA generates richer expressions, better lip-synchronization, and more natural head movements, thanks to our multi-modal motion generation network, which effectively integrates deep information across different modalities.

\subsubsection{Visualization Results in Complex Scenarios.}
We further investigate MoDA's generation performance in complex scenarios. Specifically, for the visual modality, we utilize portraits of both real humans and animated characters, each paired randomly with various audio types, including speech, singing, recitation, and others. 
\begin{table}[h]
\centering
\setlength{\tabcolsep}{7pt} 
\footnotesize
\caption{Ablation study results on major architectural choices.}
\begin{tabular}{lcccc}
\toprule
\textbf{Method}  
& \textbf{FID} $\downarrow$ 
& \textbf{FVD} $\downarrow$ 
& \textbf{Sync-C} $\uparrow$ 
& \textbf{Sync-D} $\downarrow$ \\ 
\midrule
w/ CABA 
& 47.365
& 232.291
& 5.331
& 8.692 \\
w/o C2F
& 44.548
& 221.631
& 5.535
& 8.465  \\
w/ MAF
& 48.358
& 216.982
& 5.511
& 8.527  \\
Full Model      
& 44.071
& 205.442
& 5.878
& 8.135 \\
\bottomrule
\end{tabular}
\label{tab:comparison3}
\end{table}
As shown in Fig.~\ref{fig:shiyantu2}, our method demonstrates strong performance in various complex scenarios. In addition, it is capable of efficient long-duration inference and fine-grained facial expression control.
\subsection{Ablation Study}
For more ablation experiments and implementation details, please refer to the supplementary materials.
\subsubsection{Cross-attention}\label{sec:4.3.1}
\begin{figure}
    \centering
    \includegraphics[width=1.0\linewidth]{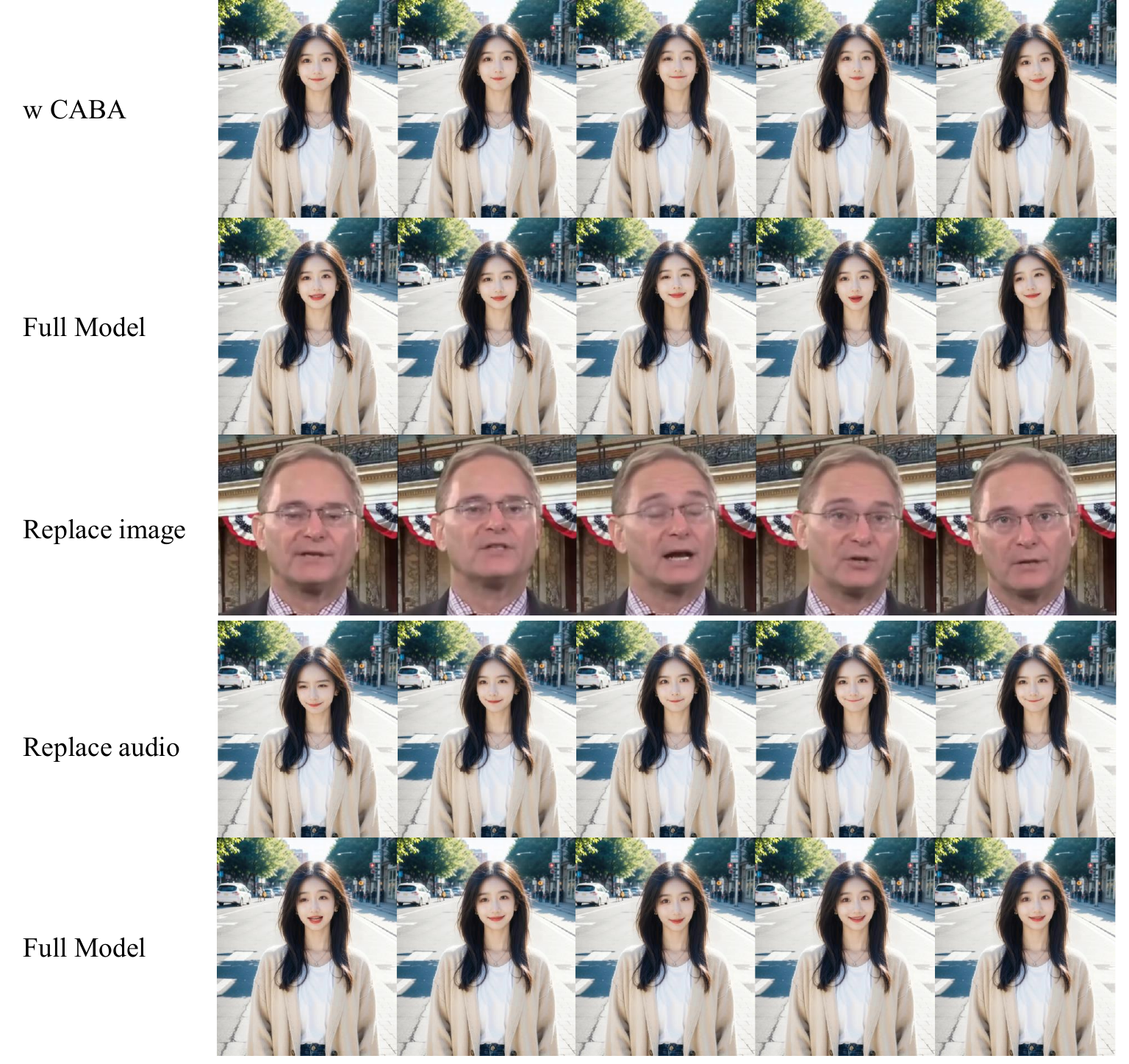}
    \caption{Visualization of ablation studies on key components.}
    \label{fig:ablation}
\end{figure}

We replace the MoDA with a Cross-Attention-Based Architecture (CABA) to perform multi-modal fusion for evaluation purposes. As shown in Table~\ref{tab:comparison3}, this substitution results in a performance drop in all metrics. Moreover, as illustrated in Fig.~\ref{fig:ablation}, the subject exhibits mouth-closing failures when lacking deep multi-modal interaction. To investigate this, we conducted two experiments under the same settings used in the w/ CABA variant as show in Fig.~\ref{fig:ablation}:
1) replacing the audio with the same image did not resolve the mouth closure issue (Replace audio);
2) replacing the image with the same audio addressed the issue (Replace image).
These results suggest that cross-attention fails to adapt audio features to different identity conditions. In contrast, MoDA enables dynamic adaptation of audio features based on contextual factors such as motion, identity, and emotion.

\subsubsection{Coarse-to-Fined Feature Fusion}
We replace the Coarse-to-Fine Feature Fusion with a fully four-stream architecture (w/o C2F). As shown in Table~\ref{tab:comparison3}, this design leads to performance degradation, particularly in synchronization and motion quality, emphasizing the necessity of progressive feature fusion for effective cross-modal integration. These results indicate that independent learning of weights introduces substantial redundant parameters, resulting in convergence challenges (with 904M parameters in the w/o C2F variant). In contrast, by gradually sharing weights, the proposed coarse-to-fine strategy significantly reduces the parameter count to 370M, leading to both greater efficiency and improved performance.

To further investigate the effectiveness of the fusion strategy, we modified the original dual-stream design by creating an MAF variant that first merges motion with the auxiliary cues (emotion + identity) into one branch, while audio stays in the other. Table~\ref{tab:comparison3} also shows that MAF performs worse than even the w/o C2F baseline. Directly fusing the heterogeneous noisy motion and auxiliary condition features confuses the network. Our full model first merges audio with the auxiliary cues; speech audio inherently carries emotion and identity signals, providing a natural bridge and delivering better accuracy and efficiency.
\subsection{Discussion}
For the disentanglement stage, we can use alternatives such as MegaPortraits or Face Vid2Vid, but like LivePortrait they degrade on large pose changes or when the subject wears complex head accessories. This may be due to limitations in the training data or intrinsic constraints of the GAN-based architecture. To address these challenges, we plan to improve the training dataset by incorporating a wider range of variations in head poses and complex head accessories. Furthermore, we will investigate more advanced video generation models~\cite{kong2024hunyuanvideo, liu2024sora} to improve the quality and expressiveness of the generated results.
\section{Conclusion}
We propose MoDA, a two-stage multi-modal diffusion framework for one-shot talking head generation. MoDA effectively leverages multi-modal information to map audio to motion sequences in an identity-agnostic latent space, which are then translated into video frames by a pre-trained face renderer. From a single portrait, MoDA produces high-quality, expressive, and controllable talking head videos, surpassing previous methods in quality, diversity, and naturalness with high efficiency.
\bibliographystyle{plain}
\bibliography{reference}
\end{document}